# Self-Assembled Telecom Color Centers in Silicon and Their Growth Environment

Jacqueline Marböck*, Enrique Prado Navarrete, Merve Karaman, Oliver E. Lang, Thomas Fromherz, Maciej O. Liedke, Andreas Wagner, Moritz Brehm*, Johannes Aberl

J. Marböck, E. Prado Navarrete, M. Karaman, O. E. Lang, T. Fromherz, M. Brehm, J. Aberl
Institute of Semiconductor and Solid-State Physics, Johannes Kepler University, Altenberger Straße 69, Linz 4040, Austria
E-mail: moritz.brehm@jku.at; jacqueline.marboeck@jku.at

M. O. Liedke, A. Wagner
Helmholtz-Zentrum Dresden-Rossendorf e.V., Institute of Radiation Physics, Dresden, 01328, Germany

**Funding:** Austrian Science Fund (FWF) [10.55776/Y1238], [10.55776/P36608], [10.55776/COE1] and the European Union - NextGenerationEU. Initiative and Networking Fund of the Helmholtz Association (FKZ VH-VI-442 Memriox), and the Helmholtz Energy Materials Characterization Platform (03ET7015)

**Abstract text.**
Artificial atoms based on color centers in silicon (SiCCs) have recently emerged as promising candidates for highly integrable and scalable key components in photonic quantum technology, including telecom single-photon sources and spin memory devices. A novel all-epitaxial fabrication technique for SiCCs, based on ultra-low-temperature (ULT) molecular beam epitaxy (MBE), addresses limitations of conventional fabrication via ion implantation, such as vertical ion straggle and collateral crystal lattice damage. This method solely relies on self-assembly of SiCCs during kinetically-limited growth of (carbon-doped) Si(:C) at ULTs ≲350°C. The latter requires an extraordinary pristine growth environment to prevent unintended defect formation caused by the incorporation of impurities from the background vapor; however, so far, no study has specifically addressed how exactly the vacuum conditions during epitaxy influence SiCC formation, their optical properties, and the quality of the surrounding crystal matrix. Here, we investigate the impact of the growth pressure and the substrate temperature on the self-assembly and photoluminescence (PL) properties of important SiCCs, such as W, G, G', and T centers. Further, we use PL and Doppler broadening variable energy positron annihilation spectroscopy to emphasize the role of the growth pressure in suppressing the luminescence background, which is crucial for advancing quantum photonics applications.

## 1. Introduction

After being long studied as detrimental radiation defects in silicon (Si) electronics [1], Si-based color centers (SiCCs) [2] have recently re-emerged due to their great potential for application as single-photon emitters [3,4] and spin-photon interfaces [5,6,7], both crucial building blocks for quantum communication and distributed quantum computing [8,9]. Thereby, a major advantage of SiCCs *vs*. competing systems [10] originates from the inherent compatibility with the mature Si photonics platform [11,12,13], the unrivaled device integration- [14] and respective tuning possibilities [15] as well as their light emission below the energy band gap of Si and thus mostly within the optical wavelength bands used for fiber-based telecommunication. Conventionally, SiCCs are created via single- [16] or double-step [17,18] ion implantation processes involving subsequent and/or intermediate thermal annealing to cure the induced damage to the crystal lattice partially, tune the predominant emitter type [19], and/or their spatial densities. However, in addition to the inevitable remaining collateral lattice damage, the resulting vertical emitter distribution strongly depends on the implanted species [20] and on the chosen implantation parameters and will commonly span tens or even hundreds of nanometers [21].

Considering the resulting drawbacks for efficient photonic- and device-integration, recently a new fabrication scheme for SiCCs has been introduced based on epitaxial self-assembly using molecular beam epitaxy (MBE) at ultra-low temperatures (ULTs) ≲350°C for Si growth [22]. In this method, SiCC formation is promoted by kinetically limiting the surface mobility of the ad-atoms during the epitaxy of Si or carbon-doped Si (Si:C) layers. Careful control of the growth temperature and doping enables the deterministic formation of different types of color centers (W, G, G', and T-center) confined to a sub-nm thick layer that can be placed at arbitrary depths below the sample surface, surrounded by a high-quality Si crystal matrix grown at slightly higher temperatures [22]. This self-assembly approach is hence particularly well suited for high-yield high-Q photonic integration [23] or for directly embedding the SiCCs into vertical p-i-n diodes [24] for electrical driving or -tuning. The confinement of SiCCs to nm-thin layers can, *i.a.,* further ensure a highly efficient use of precious, isotopically-purified $^{28}$Si material [25] as the emitters/qubits can be precisely surrounded by isotopically-tailored heterostructures, something virtually impossible for color centers created through ion implantation. Thus, the fabrication principle of self-assembled SiCCs through ULT-epitaxy with strict control of the emitter position in designed layers resembles the fabrication of some of the most advanced single-photon emitters, such as GaAs/AlGaAs quantum dots [26,27], but additionally benefits fully from the Si host platform.

However, ultra-low growth temperatures are not only required for self-assembled SiCC formation, but also for their overgrowth, as emitters like G-centers already start to dissolve at temperatures of around 250°C for the respective time scales [19]. Thus, the quality of the Si matrix surrounding the SiCCs is correlated to the quality of the ULT growth. During MBE, the growth pressure in the vacuum chamber and the rate of molecule impingement on the samples [28] are intimately linked. At typical growth temperatures for Si (>500°C) this relation is mostly irrelevant as sticking coefficients are small and unwanted molecules can almost entirely desorb from a Si surface [29]. However, at ULT (≲350°C), the incorporation of residual molecules that constitute the chamber pressure can lead to defective growth. In the previous works [22,24], the actual chamber pressure during the growth was extraordinarily low (< $10^{-10}$ mbar), leading to clean luminescence spectra, with low or no background luminescence that is typical for open volume defects in the crystal matrix.

Here, we investigate how the MBE chamber pressure during growth of the nanometer-thin active layers influences the spectral quality of self-assembled SiCCs, such as G-, G'-, W- and T-centers. We compare the PL emission from SiCCs to reference Si samples to analyze the influence of parasitic background radiation and non-radiative recombination resulting from imperfect crystal lattices. We further extend this investigation to Si samples grown for Doppler broadening variable energy positron annihilation spectroscopy (DB-VEPAS) [30,31] for a quantitative assessment of the defect density in the Si matrix at different growth temperatures.

## 2. Results and Discussion

The growth of each investigated sample started with a crystalline, high-temperature Si buffer on a bulk Si(001) substrate, followed by the growth of a 9 nm thick Si or Si:C active layer at a substrate temperature $T_G$ = 200°C. The active layer was finally overgrown with a cap at a constant temperature of either 200°C, 250°C, 300°C, or 310°C, effectively providing a post-growth annealing for the active layer. However, as some SiCCs start to dissolve at such low temperatures [1,22], the choice of a particular capping layer temperature is always a tradeoff between SiCC concentration and high-quality crystalline Si overgrowth. **Figure 1a** shows the respective sample structure for the carbon-doped Si:C emitter layer (c = 3.8×10$^{19}$ cm$^{-3}$) and Figure 1b for the series with undoped Si instead of the Si:C emitter layer. To account for possible unintended SiCC formation within the Si capping layer at relatively low capping temperatures, corresponding reference samples have been grown without the active emitter layer to isolate the contribution of the Si capping layer on the overall luminescence signal. The respective sample structure is pictured in Figure 1c.

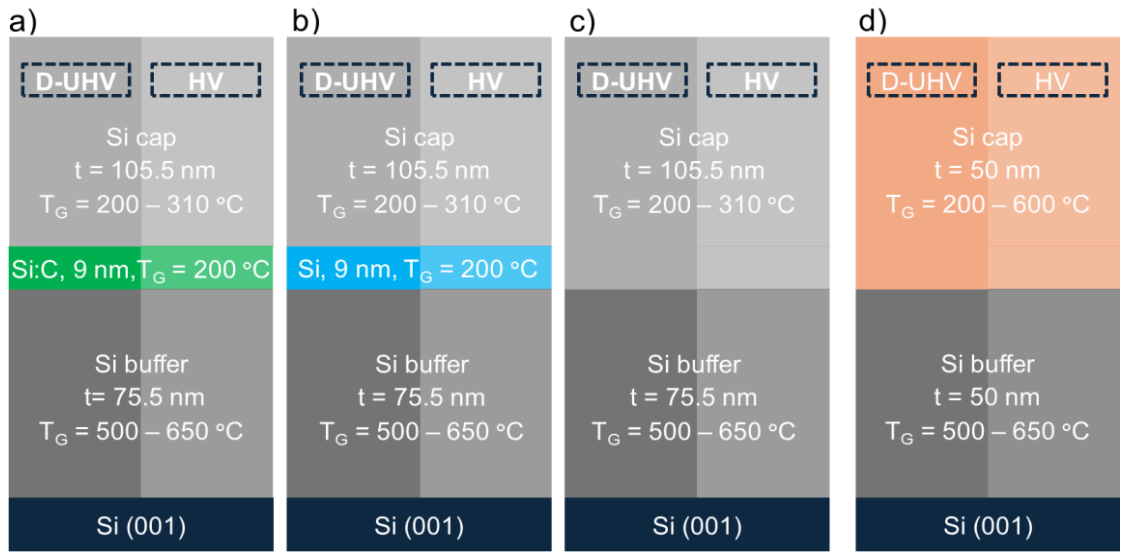

**Figure 1.** Sample schemes for SiCCs grown via ULT MBE, either under D-UHV or HV conditions. The SiCCs are formed through kinetically limited growth. (a), (b) A 9 nm thick active emitter layer, deposited at growth temperatures $T_G$ = 200°C, is embedded between a highly crystalline Si buffer layer and a Si capping layer grown at varying temperatures $T_{cap}$ = 200°C, 250°C, 300°C, and 310°C. The active layer in sample series (a) is carbon-doped ULT-Si (c = 3.8×10$^{19}$ cm$^{-3}$), whereas series (b) comprises undoped, ULT-grown Si. Series (c) serves as a reference for (a) and (b) to identify the influence of $T_{cap}$ on the color center formation. For sample series (d), a 50 nm thick Si layer was grown at varying $T_G$ = 200°C, 300°C, 350°C, 600°C on a high-temperature Si buffer layer. These samples were employed in PL and DB-VEPAS measurements.

One additional sample series, sample structure Figure 1d, was intended to study the influence of the Si growth temperature on crystal damage formation in a wider temperature window, up to $T_G = 600°C$, by means of photoluminescence (PL) and Doppler broadening variable energy positron annihilation spectroscopy (DB-VEPAS). All of these sample series were repeated and grown under different vacuum conditions, i.e., under high-vacuum (HV) conditions with growth pressures in the range from $2\times10^{-8}$ mbar to $2\times10^{-7}$ mbar and under deep-ultra-high-vacuum conditions (D-UHV) for which the active layers were grown at chamber pressures between $\sim3\times10^{-10}$ mbar, and below the detection limit of the pressure gauge ($< 5\times10^{-11}$ mbar).

## 2.1. G'- and G-centers

**Figure 2** depicts the semi-logarithmic low-temperature PL spectra of samples with structures according to Figures 1a and 1c that were each epitaxially grown at two fundamentally different growth pressures. The columns in the matrix in Figure 2 refer to different growth temperatures of the Si capping layer ($T_{cap}$), further represented using different colors for $T_{cap} = 200°C$ (black), 250°C (red), 300°C (blue), and 310°C (green). The topmost row (Figures 2a-d) shows results from samples, dominated by G'-center emission (cf. structure in Figure 1a), for which 9 nm of Si:C were deposited at 200°C, followed by a 105.5 nm thick Si capping layer, grown at the respective $T_{cap}$. Darker and lighter colored spectra denote results from samples grown under D-UHV conditions, and HV conditions, respectively, but under otherwise same growth conditions. For $T_{cap} = 200°C$ and D-UHV conditions (black spectrum in Figure 1a), a weak PL intensity signal from bulk Si at 1128 nm is accompanied by characteristic zero-phonon line (ZPL) emission from W-centers at ~1218 nm and its respective phonon side band (PSB), together with ZPLs from G-centers, emitting at ~1278 nm and G'-centers at 1300 nm, and the respective PSB and local phonon modes [1,22,32]. The G'-center is a defect of the G-center family, consisting of a conventional G-center (two substitutional C-atoms linked via one interstitial Si atom) and an additional C-atom at a nearest neighbor substitutional lattice site [22]. This color center is predominant in ULT Si:C layers since the C-evaporation source emits mainly $C_3$-molecules and the ULT conditions limit the dissociation of the $C_3$-clusters [22].

The spectral shape is drastically different when the growth is performed at worse chamber pressures (grey spectrum in Figure 2a). Only a faint background signal with a very weak sign for G'-center ZPL at 1300 nm is visible, indicating very clearly that non-radiative recombination processes dominate the energy relaxation of the excited electron-hole pairs. The faint, broad background, rising in PL intensity from 1300 nm to the detector cut-off at 1550 nm is typically related to the open volume defects within the Si crystal [33].

This behavior changes if at HV the Si:C layer is overgrown with Si, deposited at 250°C (Figure 2b, light red spectrum). Besides a PL signal from Si bulk, a clear ZPL from G'-centers and an associated phonon side band is visible. However, we emphasize that the sole presence of G'-center emission is not an adequate parameter to judge the sample quality, as is immediately evident by comparing the light and dark red spectra in Figure 2b. The latter corresponds to the same sample structure ($T_{cap}$ = 250°C) but the whole deposition process was performed in D-UHV.

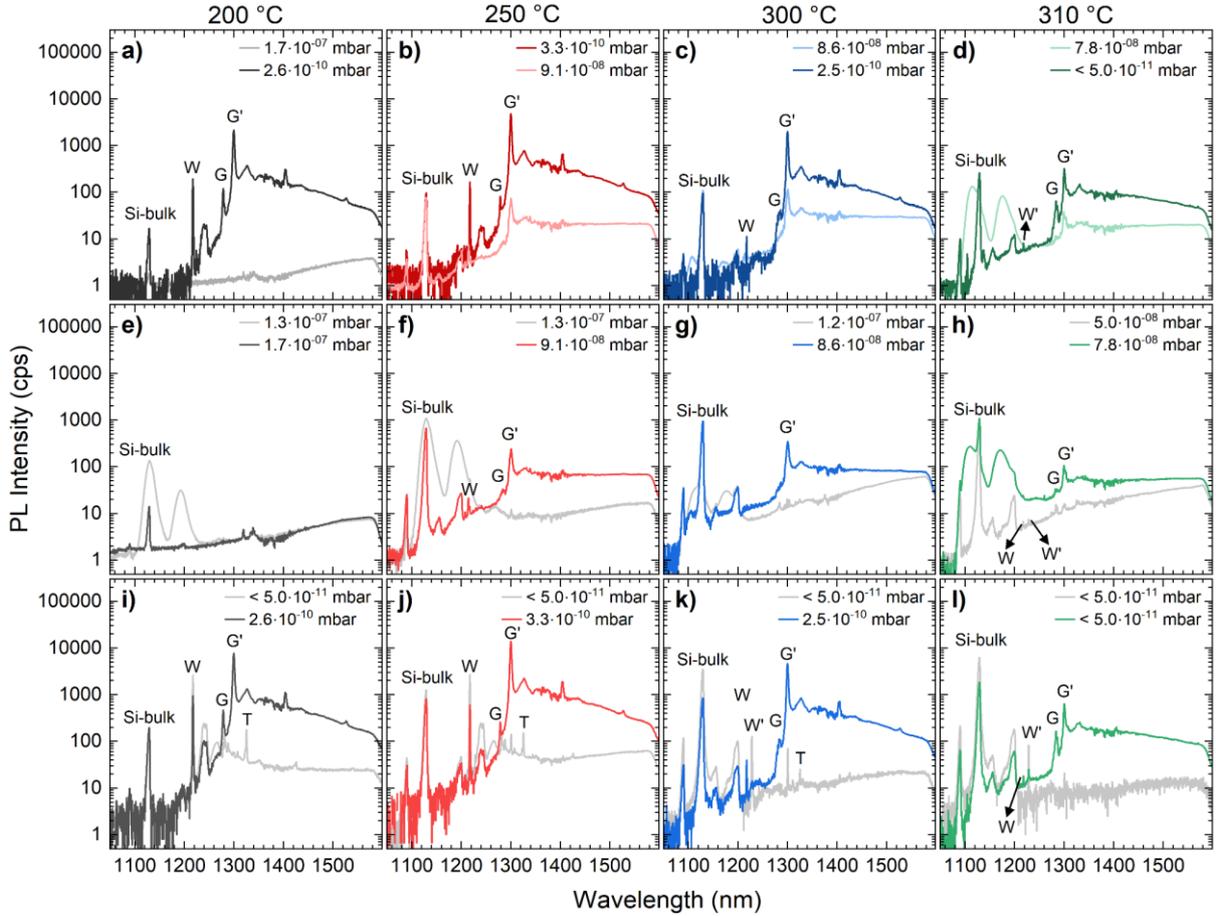

**Figure 2:** Low-temperature PL measurements performed at 15 K of fabricated SiCC samples, grown at different pressures and overgrown at different $T_{cap}$. (a) - (d) PL spectra for two identically grown samples except for the background pressure during growth $p_g$ in the MBE chamber, ultra-high vacuum (UHV) (dark coloring) and high vacuum (HV) (light coloring) for different cap temperatures $T_{cap}$ = 200°C (black), 250°C (red), 300°C (blue), 310°C (green). The sample excitation power was 0.195 mW. (e) - (h) The samples are compared to the corresponding references (grey curves) deposited in HV and (i) - (l) in D-UHV. All spectra in (e)-(l) were recorded for an excitation power of 0.974 mW.

The overall PL intensity from G'-centers of the HV-sample is reduced by about two orders of magnitude as compared to the D-UHV sample. Moreover, we note that only the D-UHV sample exhibits pronounced W-center emission that originates from kinetically limited growth of a pure Si capping layer deposited at 250°C (see discussion of Figure 2j below), leading to the W-center's tri-interstitial [1,34] formation. Instead, under higher growth pressure, the incorporation of impurity atoms inhibits W-center emission, and, thus, the absence of W-center emission at low $T_{cap}$ and HV-conditions should not be seen as a sign for high-quality crystalline growth, but rather vice versa.

A similar trend between D-UHV and HV samples is observed if the Si:C layer is overgrown at $T_{cap}$ = 300 °C (Figure 2c). There, the difference in G'-center's PL emission intensity is still very evident (~factor 10), but slightly less pronounced than at lower $T_{cap}$. This behavior is preserved for $T_{cap}$ = 310°C (Figure 2d), however, $T_{cap}$ = 310°C already leads to a decrease in the G'-center emission which is associated with the thermal annihilation of the emitters. Nevertheless, for the sample grown under D-UHV conditions, the PL emission intensity from the G'-centers ZPL is increased by about than an order of magnitude as compared to the growth at HV conditions. Furthermore, the incorporation of point defects at HV conditions leads to a broadening of the Si-related PL and a broadening of the full width at half maximum (FWHM) of the G'-center ensemble ZPL (3.25 nm for D-UHV and 4.25 nm for HV).

Figures 2e-h compare the PL spectra of the Si:C layer containing samples grown at HV with the respective reference samples for which the Si:C layer was replaced by a Si layer, grown at the respective $T_{cap}$. These reference samples were grown under similar HV conditions. For $T_{cap}$ = 200°C, the spectral shape below the bandgap of Si is dominated for both the Si:C and the reference sample by the broad defect-related PL band that increases in PL intensity towards the detector cut off at 1580 nm.

The spectra of the HV-reference samples (grey spectra in Figures 2e-h) exhibit no or only weak traces of W-, G-, G'-, and T-center emission. The comparison of these reference spectra to the ones containing the Si:C layer (black, red, blue, and green spectra in Figures 2e-h) reveals two interesting points. First, the PL emission intensity of the HV samples containing the Si:C layer is almost constant in the spectral range from 1350 nm to 1580 nm. This spectral shape seems to be a consequence of the sum of PL intensity contributions from the PSB of the G'-centers, which is decreasing with increasing wavelength, and the defect-related PL band, which is increasing in intensity at higher wavelengths. Second, the intensity of the defect-related PL band is increasing with increasing $T_{cap}$ from 200°C to 300°C. Thus, in this case, and somewhat counter-intuitively, a lower PL intensity of the defect-band is not a sign for better sample quality, but rather the consequence of the enhanced incorporation of defects at lower $T_{cap}$s that lead to non-radiative recombination rather than contributing to the defect PL band.

This contrasts with the behavior when the reference samples were grown at D-UHV conditions (grey spectra in Figures 2i-l). Increasing $T_{cap}$ from 250°C to 310°C leads to a decrease of the intensity of the defect-related PL background, indicating the highest Si matrix quality at $T_{cap}$ = 310°C. However, for such overgrowth conditions, thermal dissolution of the G'-centers is already present, as indicated by a factor 7 reduced G'-center ZPL for $T_{cap}$ = 310°C as compared to $T_{cap}$ = 300°C.

Furthermore, for D-UHV conditions (Figures 2i-l) and the various $T_{cap}$s, the W-center PL emission intensity is higher for the Si reference samples than for the Si:C containing samples, implying that the W-center emission originates predominantly from the Si capping layer. Focusing further on the W-center for the D-UHV samples, a peculiar behavior can be observed as $T_{cap}$ is increased to 300°C and above (Figures 2k,l). At $T_{cap}$ = 300°C, for the reference sample (grey spectrum in Figure 2k), no W-center ZPL emission exists at 1218 nm; instead, a pronounced, sharp emission line at 1228 nm can be seen which is also present in the reference spectrum for $T_{cap}$ = 310°C. The presence of both ZPL lines, 1218 nm and 1228 nm that are similar in FWHM, (FWHM = 1.19 nm for 1218 nm and FWHM = 1.60 nm for 1228), in the green spectrum in Figure 2l suggests that the emission center found here with a ZPL at 1228 nm is likely associated with the W-center, a derivate that we here name W'-center, that forms under slightly higher thermal budget. Note that many additional color centers have been reported that emerge after annealing of W-center samples, but these centers, like the X and J centers [19,35,36] emit at shorter wavelengths than the W center, while the W' center emits at longer wavelengths. The identification of the atomic origin of the W'-centers is beyond the scope of this work.

**2.2. T-centers**

T-centers, hydrogen-(H) 2C point defects in Si, are versatile for quantum applications due to their favorable spin properties like long coherence times, and optically addressable spin states, enabling the creation of a spin-photon interface [5,7,25,37]. Here, we demonstrate T-center formation via self-assembly, employing the sample scheme shown in Figure 1b. Exemplary, the green spectra in **Figure 3** depict the spectroscopic fingerprints from self-assembled T-centers for which the 9 nm thick active layer was grown at 200°C and $T_{cap}$ was 310°C. The spectra in Figures 3a-c originate from samples that were grown at different growth pressures, but otherwise identical conditions. The grey spectra originate from reference samples, for which the ULT-layers were grown at $T_{cap}$ = 310°C throughout.

By comparing the green spectra in Figures 3a,b, we study the influence of different growth pressure constituents on the resulting color center emission. While the total growth pressures during the 9 nm thick ULT Si layer are comparable (~$5\times10^{-8}$ mbar versus $3\times10^{-8}$ mbar), the PL spectra are very different, i.e., the T-center ZPL in Figure 3b is increased 23-fold as compared to Figure 3a. The overall pressures are dominated by $H_2$, while the partial pressures of other residual gases were different, e.g., the concentrations of C, $CH_4$, and $CO_2$ were at least a factor of 10 higher for Figure 3a than for Figure 3b (C: $1.9\times10^{-12}$ mbar vs. $1.1\times10^{-13}$ mbar, $CH_4$: $7.5\times10^{-12}$ mbar vs. $6\times10^{-13}$ mbar, $CO_2$: $1.9\times10^{-12}$ mbar vs. $1.9\times10^{-13}$ mbar) during the growth of the active layer, as verified by mass spectrometer analysis.

Figure 3c verifies that the self-assembled T Centers (green curve) are fully confined within the 9 nm-thin layers, forming during $T_G$ = 200°C, while the increased thermal budget of the Si capping layer growth prevents T-center formation during overgrowth. In the grey reference signal in Figure 3c, no sign of T-center emission is present, and the overall PL count rate is low, implying a good Si matrix quality and T-center formation solely in the ULT-Si layer.

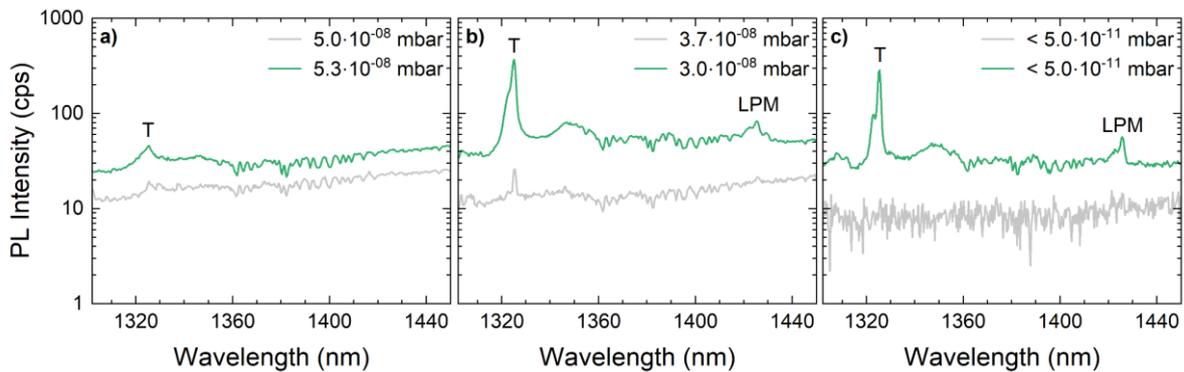

**Figure 3:** Influence of MBE growth pressure and pressure constituents on the PL intensity of T-centers. Note the semilogarithmic PL intensity scale. 9 nm thick Si layers were grown at $T_G$ = 200°C and overgrown at $T_{cap}$ = 310°C (green curves). For the reference samples (grey spectra) the 9 nm thick layers and the Si capping layer were grown at $T_{cap}$ = 310°C. The samples were grown at different growth pressures, ranging from a) and b) HV to c) UHV.

The reference samples grown at higher growth pressure (Figures 3a,b) exhibit, however, T-Center emission, which is a direct result of the increased molecular density containing C and H that are incorporated at $T_{cap}$ = 310°C. We note that the FWHM of the ensemble T-center ZPL becomes narrower with lower pressure (HV 2.2 nm and 1.5 nm for D-UHV), further pointing to the influence of the matrix quality on the color center emission. Furthermore, a clear trend of lowered background PL with improved vacuum conditions can be observed by comparing the reference spectra in Figure 3.

### 2.3. W-centers

In contrast to carbon-related Si color centers, such as T-centers, G-centers, and C-center families, W-centers are known to be self-interstitials in Si [19,34,35,36]. Similar to the previously discussed self-assembled T-, and G'- and G-centers, the formation of W-centers using epitaxy is also influenced by the growth pressure and growth pressure constituents. W-centers can be preferentially created by following the sample growth scheme depicted in Figure 1b. Here, 9 nm of ULT Si are grown at 200°C, followed by the growth of a Si capping layer at $T_{cap}$ = 250°C. The relatively low $T_{cap}$ limits W-center dissolution during the Si capping layer growth. However, in contrast to G'-centers and T-centers, for W centers, a perfect selectivity for positioning within the 9 nm thick layer is not achieved (see **Supporting Information**). For the highest investigated growth pressure of ~5×10$^{-8}$ mbar, for which the partial pressures of C-related molecules are still in the 10$^{-12}$ mbar range, the carrier relaxation in PL is dominated by non-radiative pathways, and no pronounced signal from W-centers is observed, see Supporting information, Figure S2a. For lowered partial pressures of C-related molecules (Figure S2b), a W-center ZPL at 1218 nm, a W'-center ZPL at 1228 nm and the typical PSB are visible. However, strong W-center and W'-center emission is also present in the reference sample, which supports the claim that position selectivity is not achieved for the W-centers. While more than one-third of the W-center emission seems to originate from the cap, the W'-center ZPL in the grey spectrum in Figure S2b suggests that the W'-centers in the red spectrum are almost entirely situated in the cap.

### 2.4. Photoluminescence background and Si matrix quality

The quality of the matrix surrounding quantum systems is crucial for potential applications. Therefore, we investigated the influence of the growth pressure on the quality of the Si matrix in a wider $T_G$ window and added a quantitative measure of point defect densities versus $T_G$ using DB-VEPAS. Therefore, in **Figure 4**, we investigate the sample series described in Figure 1d, consisting of a high-quality Si buffer layer, followed by a 50 nm thick Si layer grown at temperatures $T_G$ = 200°C (black spectra in Figure 4), 300°C (blue), 350°C (violet), and 600°C (orange), respectively. The samples were grown at different growth pressures, ranging from HV (light coloring) to D-UHV (dark coloring) in the spectra in Figure 4, plotted on a semi-logarithmic scale. Each spectrum features the characteristic Si bulk replica in the wavelength range from ~1100 nm to ~1200 nm.

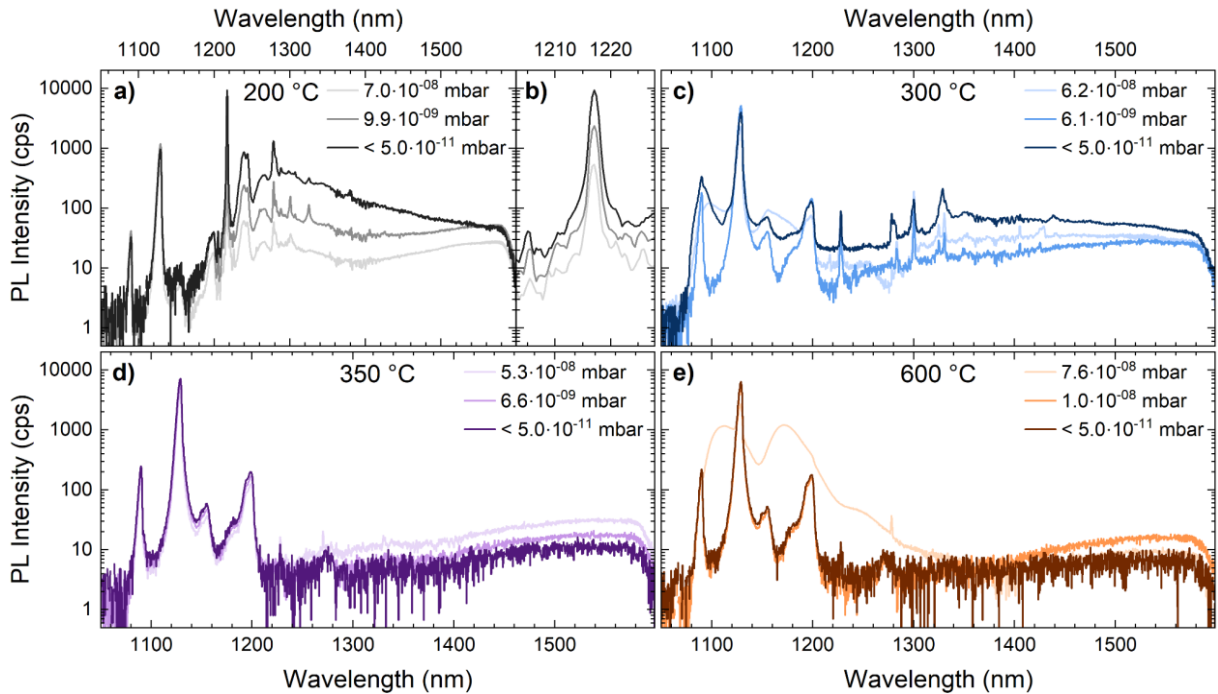

**Figure 4.** PL spectra of Si deposited at various growth pressures and temperatures. A 50 nm thick Si buffer was grown at 500 - 650°C, followed by 50 nm of Si grown at different temperatures $T_G$ = 200°C (black spectra), 300°C (blue), 350°C (violet) and 600°C (orange spectra). Each structure was grown at different vacuum conditions ranging from ~7×10$^{-8}$ mbar to 10$^{-11}$ mbar (light to dark coloring of the spectra).

Figure 4a depicts the spectra for $T_G$ = 200°C. Below the Si band gap, the PL spectra are dominated by emission from W-centers with a ZPL at 1218 nm and the typical PSB. For higher growth pressures, the overall spectral shape remains; however, the emission is lowered by more than one order of magnitude due to parasitic non-radiative recombination, see close-up of the ZPL of the W-centers in Figure 4b. In Figure 4a, there are, however, subtle differences between the spectra. For the highest pressures, the spectral shapes of the PSB are distorted by the PL defect band, and pronounced G'-center ZPL emission is visible at 1300 nm.

Epitaxial growth of Si at $T_G$ = 300°C still leads to the formation of color centers. However, in Figure 4c, it can be seen that the SiCC emission below the Si band gap is about two orders of magnitude lower than for the PL emission from Si bulk. The increased thermal budget at $T_G$ = 300°C, compared to 200°C enhances the surface kinetics of the Si adatoms during growth, leading to fewer interstitial defects and, thus, the PL intensity of the W-center PL emission is drastically reduced. Notably, as previously observed in Figure 2, the W-center emission peak shifts to 1228 nm, the W'-center ZPL position, while for the highest pressure, a faint peak at 1218 nm remains. Additional PL peaks related to the G-centers, G'centers, and T-centers are present, as carbon can still be captured during the long growth interrupt (12 min 30 s) between the high-temperature Si buffer layer and the ULT Si layer, and the ULT layer growth itself. Hydrogen and hydrocarbon species etc., impinging on the Si surface begin to desorb at temperatures of ~350°C. This explains the pronounced change in spectral shape of Si, grown at $T_G = 300°C$ (Figure 4c) and $T_G = 350°C$ (Figure 4d). For the latter, only background PL is visible for which the PL intensity is decreasing with decreasing growth pressure. For the highest growth pressure (~$5 \times 10^{-8}$ mbar), weak signals from the W'-center, G-center and T-center could be observed (Figure 4d). We compare these results of ULT Si growth to conventional, high-temperature Si epitaxy ($T_G = 600°C$) in Figure 4e. Several observations can be made. First, the background PL signal at the lower growth pressures is comparable to the results at $T_G = 350°C$, directly indicating that even for ULT-Si epitaxy, an excellent crystalline quality can be achieved. If the PL spectra for $T_G = 350°C$ and 600°C are plotted on a linear scale, only a flat baseline would be visible (see Supporting Information, Figure S1). Second, there is virtually no difference between the spectral shapes for the samples grown at D-UHV and at $1 \times 10^{-8}$ mbar, for which the pressure is dominated by hydrogen molecules. Note that this is not the case for $T_G = 350°C$, where a difference in the spectra can be seen. This indicates that high temperature growth is much more forgiving concerning the growth environment and that special care concerning excellent growth pressures needs to be taken for ULT growth. Third, even if the growth temperature is high, defect incorporation seems to occur if the partial pressures of C, $CH_4$, and $CO_2$ are high, as can be seen by the distorted Si bulk PL signal and the increased defect band PL intensity around 1500 nm.

## 2.5. Doppler broadening variable energy positron annihilation spectroscopy (DB-VEPAS)

Although PL measurements suggest that samples grown at D-UHV conditions and at $T_G = 350°C$ and 600°C are basically defect-free, we intend to confirm and quantify our findings using Doppler broadening variable energy positron annihilation spectroscopy (DB-VEPAS). Details regarding positron physics and methodology can be found in the Experimental Section. The results are presented in **Figure 5**. The parameter S, which indicates the fraction of positrons annihilating with low momentum; valence electrons, scales mostly with defect concentration; however, it depends on the defect size and material composition, too. The parameter W, the overlap of the positron wavefunction with high-momentum core electrons, indicates the defect's atomic surroundings.

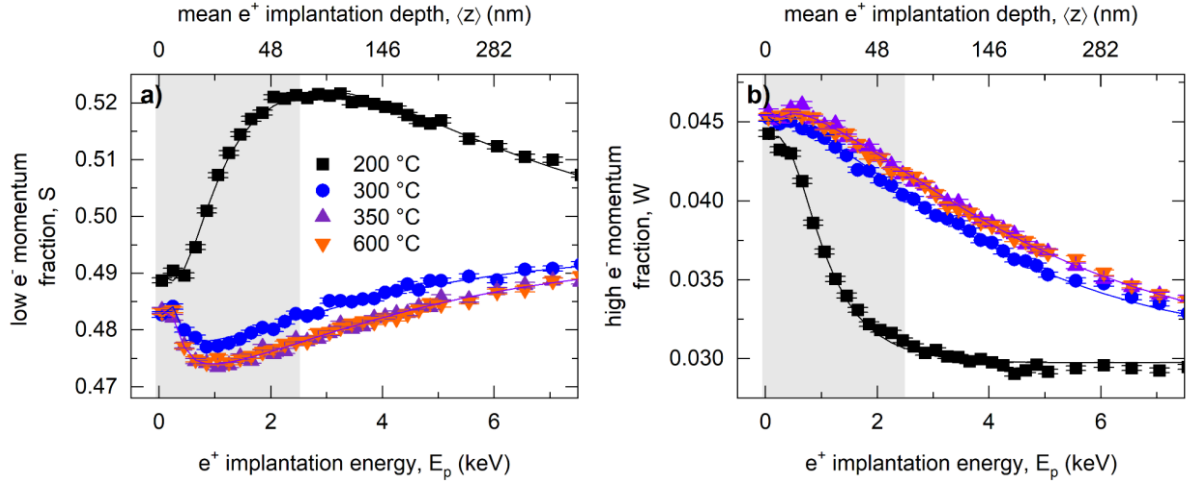

**Figure 5.** Positron annihilation spectroscopy for Si samples. (a) The S-parameter is a fraction of positrons annihilating with low-momentum valence electrons and represents vacancy-type defects and their concentration. (b) The W-parameter approximates the overlap of the positron wavefunction with high-momentum core electrons. For S at $T_G$ = 200°C sample (black squares), many positrons annihilate, the sample growth at $T_G$ = 300°C (blue circles) is almost defect-free, and for $T_G$ = 350°C and 600°C (violet and orange triangles), the defect level is below the sensitivity limit. The grey shaded area marks the thickness of the topmost epitaxial layer.

Due to the low density of Si, positrons can be implanted down to about 4 µm, using the available maximal $E_p$ = 35 keV, but the region of the epitaxial layer of interest is in the direct vicinity of the surface. A change in defect concentration is visible directly in both S and W parameters as a difference of curve slopes in the low $E_p$ range (<3-8 keV). The S and W curves are qualitatively inverted to each other for all the studied samples, which suggests no significant variations in defect surroundings. For the sub-surface region $E_p$ < 1 keV, an oxide layer is found for all the samples, represented as a plateau-like region of the curve and originating from the native Si oxide. In the case of $T_G$ = 200°C sample, the S parameter increases to a depth of about 50 nm, which is exactly the overlayer thickness. Then a short plateau of another 50 nm is visible, followed by the S-parameter decrease, where positrons start to see a defect-free substrate. The ratio $S/S_{bulk}$=1.055(2) is in the defect size range expected for a vacancy cluster of three to four agglomerated vacancies [38,39]. The results were fitted in order to obtain an effective positron diffusion length $L_+$ and to calculate the defect density [40] (see Experimental Section).

The largest defect density of ~$1.1\times10^{20}$ cm$^{-3}$ is found for $T_G = 200°C$, leading to a rapid $S(E_p)$ increase and $W(E_p)$ decrease for $E_p < 2$ keV. For $E_p > 2$ keV, the slopes change more slowly, which is a fingerprint of low defect concentration (basically defect-free within the method sensitivity). Already for $T_G = 300°C$, slopes of $S(E_p)$ and $W(E_p)$ are much lower, which indicates much lower defect density compared to $T_G = 200°C$. For $T_G = 300°C$, the defect concentration was found to be ~$6\times10^{16}$ cm$^{-3}$, while for $T_G = 350°C$, the defect density is the same as for $T_G = 600°C$ (~$1\times10^{16}$ cm$^{-3}$), which suggests a nearly defect-free material. Beyond this work on Si, D-UHV conditions and ULT for group-IV MBE growth also allow for the epitaxy of highly supersaturated strained SiGe [41], Ge [42] and GeSn layers [43] on Si of high crystalline quality that can be used for next-generation nanoelectronic devices based on reconfigurable transistors (RFETS) [44], RFET circuits [45] as well as cryogenic transistors [46]. For these ULT-grown Ge-based nanolayers, the role of the growth pressure on defect formation and the related applications remains to be investigated.

## 3. Conclusion

In summary, our investigations demonstrate that the parameter space of growth temperature, growth pressure and pressure constituents has a crucial influence on the crystallinity of Si and the formation of self-assembled SiCCs at ULTs. In ULT growth at $T_G \lesssim 350°C$, impurity atoms constituting the growth pressure stick to the surface, leading to crystal modifications that manifest themselves as color centers, defect-background PL, or non-radiative recombination centers. While for conventional Si growth temperatures, growth pressures of ~$10^{-9}$ mbar seem to be sufficient for good quality epitaxy, for ULT growth, the pressure should be as low as possible. Using kinetically-induced self-assembly, color centers, such as the G'-center or the T-center can be fabricated in well-defined nanolayers, while keeping the surrounding Si matrix of highest quality. Thus, this study highlights the potential of self-assembled color centers as building blocks for quantum photonics applications and benchmarks the role of temperature and pressure in their fabrication.

**Experimental Section**

*MBE growth*: The samples were grown on $17.5 \times 17.5$ mm$^2$ pieces of intrinsic float zone (FZ) Si (001) (resistivity $R_0 > 10000$ $\Omega$cm) using solid source molecular beam epitaxy in a Riber

SIVA45 system. The substrate cleaning procedure included pre-cleaning with solvents, ozone cleaning, Radio Corporation of America (RCA) cleaning, and a 1-minute dip in diluted hydrofluoric acid (HF, 1%) to remove the native Si oxide. The samples were degassed at 700°C for 15 min, followed by 30 min of conditioning at 310°C. First, a 75.5 nm thick Si buffer was grown while linearly ramping the temperature from 500°C – 650°C to avoid carbon segregation [47], followed by a 9 nm thick layer of either Si or Si:C grown at $T_G$ = 200°C. For the Si:C layer, a C-doping concentration of $3.8\times10^{19}$ cm$^{-3}$ was used. Finally, this active layer was capped by a 105.5 nm thick Si matrix that was grown at a growth rate of 0.5 Å/s and at different growth temperatures of $T_{cap}$ = 200°C, 250°C, 300°C, 310°C, respectively. The corresponding sample structures are pictured in **Figures** 1a and 1b. Furthermore, to identify emitters originating from the ULT-grown cap, the samples were regrown without the active emitter layer as reference samples, see the sample scheme in Figure 1c. Finally, to determine the growth temperature required for defect-free Si overgrowth, 50 nm of Si was deposited at a rate of 0.5 Å/s and at temperatures of 200°C, 300°C, 350°C, and 600°C on a high-temperature Si buffer layer, following the sample structure shown in Figure 1d. Each sample structure was grown in D-UHV and twice in HV to investigate the light emission properties of the color centers for different vacuum conditions obtained at different stages after a chamber opening. The residual gas species was in situ analyzed using a Pfeiffer QMG 250 M1 PrismaPro mass spectrometer.

*Photoluminescence measurements*: For micro-PL measurements, a part of each sample was cleaved into a 4×4 mm$^2$ piece to be glued onto the cold finger of a liquid-helium (LHe) flow cryostat. For excitation, a continuous-wave (cw) diode-pumped solid-state (DPSS) laser emitting at 473 nm was used. To further study the effects of the growth pressure, the PL spectra were recorded for two different excitation powers, 0.974 mW and 0.195 mW, measured above the cryostat window. The laser was focused using an infinity- and glass-corrected Olympus 50x microscope objective with 0.65 numerical aperture (NA) to a spot of ~2 µm diameter, and the luminescence signal was collected through the same objective. The µ-PL spectra were recorded via a 750 mm focal-length Czerny Turner spectrometer with an 85 gr/mm grating connected to a liquid-nitrogen LN$_2$ cooled 1024 pixel InGaAs line detector. The sample temperature was 15 K in all cases.

*Doppler broadening variable energy positron annihilation spectroscopy (DB-VEPAS)*: DB-VEPAS measurements have been conducted at the apparatus for in-situ defect analysis (AIDA) [48] of the slow positron beamline (SPONSOR) [49]. Positrons have been accelerated and implanted into samples in the kinetic range of $E_p$ = 0.05 – 35 keV, which realizes depth profiling. A mean positron implantation depth was approximated using a simple material density (ρ) dependent formula [50]:

$$\langle z \rangle \, [\text{nm}] = \frac{36}{\rho \, [\text{g} \cdot \text{cm}^{-3}]} E_p^{1.62} \, [\text{keV}]$$

<z> approximates the depth and cannot be treated as an absolute measure since it does not account for positron diffusion. Implanted into a material, positrons thermalize by losing their kinetic energy. Subsequently, after a diffusion, positrons will annihilate in delocalized lattice sites or localize in vacancy-like defects and their agglomerates, emitting typically at least two anti-collinear 511 keV gamma photons once they meet electrons. Since at the annihilation site thermalized positrons have negligible momentum compared to the electrons, a *broadening of the 511 keV line* is observed, which is mostly caused by the momentum of the electrons. The annihilation signals are measured with a high-purity Ge detector (energy resolution of 1.09 ± 0.01 keV at 511 keV). The spectrum broadening is characterized by two distinct parameters S and W defined as a fraction of the annihilation line in the middle (511±0.70 keV) and outer regions (508.56±0.30 keV and 513.44±0.30 keV), respectively. The S-parameter is a fraction of positrons annihilating with low-momentum valence electrons and represents a superposition of the size and concentration of empty spaces in crystals. The W-parameter approximates the overlap of the positron wavefunction with high-momentum core electrons and indicates a chemical vicinity of the annihilation place.

*Calculation of the positron diffusion length using VEPfit:* For the analysis of positron diffusion length, $L_+$, which scales inversely proportional to defect concentration, the so-called VEPFit code [40,51] has been utilized, which permits fitting $S(E_p)$ and $W(E_p)$ curves for single and multilayered systems, particularly, to acquire thickness d, $L_+$, and specific S- and W-parameters for each layer within a stack. The calculated $L_+$ and corresponding defect density $c_V$ is presented in **Table 1**. The following material densities have been utilized for Si and $SiO_2$ native oxide, respectively: $\rho_{Si}$ = 2.329 g·cm$^{-3}$ and $\rho_{SiO2}$ = 2.65 g·cm$^{-3}$. $L_+$ of $SiO_2$ was fixed to 1 nm and its thickness d to 5 nm, which fairly describes a large defect concentration in amorphous oxides [52] and is a good approximation to a typical native oxide thickness, respectively. The absolute oxide thickness is not crucial for the fit of the Si and choosing values in the range of 1-5nm does not influence the overall result significantly. The calculated S-parameter and W-parameter values of the Si-layer for different temperatures are similar for 300°C and 350°C, $S_{layer}$=0.494(1) and $W_{layer}$=0.0302(1), and substantially increase to $S_{layer}$=0.5242(2) and $W_{layer}$=0.0295(1) for T=200°C, which agrees well with the experimental observations.

| Sample | $L_+$ (nm) | $c_V \cdot 10^{16}$ (cm$^{-3}$) |
|---|---|---|
| **200°C** | 6.0±0.4 | 11000±737 |
| **300°C** | 168±2 | 6±0.4 |
| **350°C** | 210±2 | 1±0.067 |

**Table 1.** Effective diffusion length $L_+$ and defect concentration $c_V$, calculated using VEPFit code.

*Calculation of defect concentration based on positron diffusion length:* Assuming that the films mostly contain only one major defect type the vacancy concentration $c_V$ can be calculated as follows [53]:

$$c_V = \frac{N}{\nu_V \tau_B}\left(\frac{L_{+,B}^2}{L_+^2} - 1\right),$$

where $N = 5\times10^{22}$ at/cm$^3$ is an atomic density of Si, $\nu_V \approx 3\cdot10^{15}$ s$^{-1}$ is a specific positron trapping rate (trapping coefficient) in a negative vacancy [54], $\tau_B \approx 218$ ps [55] and $L_{+,B}$ =220 nm [56] are a bulk lifetime and a diffusion length in a defect-free material. The results are presented in Table 1, concluding that the sample grown at 350°C is effectively defect-free ($L_+ \approx 210$ nm), whereas the sample grown at 300°C exhibits a very small concentration of defects.

**Acknowledgements**

This research was funded in whole or in part by the Austrian Science Fund (FWF) [10.55776/Y1238], [10.55776/P36608], [10.55776/COE1] and the European Union - NextGenerationEU. For open access purposes, the author has applied a CC BY public copyright license to any author-accepted manuscript version arising from this submission. This work was partially supported by the Initiative and Networking Fund of the Helmholtz Association (FKZ VH-VI-442 Memriox), and the Helmholtz Energy Materials Characterization Platform (03ET7015). Parts of this research were carried out at ELBE at the Helmholtz-Zentrum Dresden - Rossendorf e.V., a member of the Helmholtz Association. We would like to thank the facility staff for their assistance.

**Self-Assembled Telecom Color Centers in Silicon and their Growth Environment**

Jacqueline Marböck*, Enrique Prado Navarrete, Merve Karaman, Oliver E. Lang, Thomas Fromherz, Maciej O. Liedke, Andreas Wagner, Moritz Brehm*, Johannes Aberl

**Photoluminescence spectra of Si grown at $T_G = 350°C$ and $600°C$**

In Figure S1, the PL spectra of the sample series, described in Figure 1(d) of the main text, are plotted on a linear and semilogarithmic scale for two representative growth temperatures $T_G = 350°C$ and $T_G = 600°C$. In the main text, we have discussed the role of the background pressure on the defect PL band, which is a crucial parameter to control towards quantum photonics applications based on Si color centers fabricated via ULT-MBE. Here, we emphasize that the quality differences of the matrix can only be seen if the PL intensity is plotted on a semilogarithmic scale. Figure S1 demonstrates that on a linear intensity scale, as commonly plotted for Si-based color centers [3]. Besides bright features due to bulk Si, only flat baselines are visible in PL intensity below the Si bandgap, and no differences in the quality of the matrix are observable.

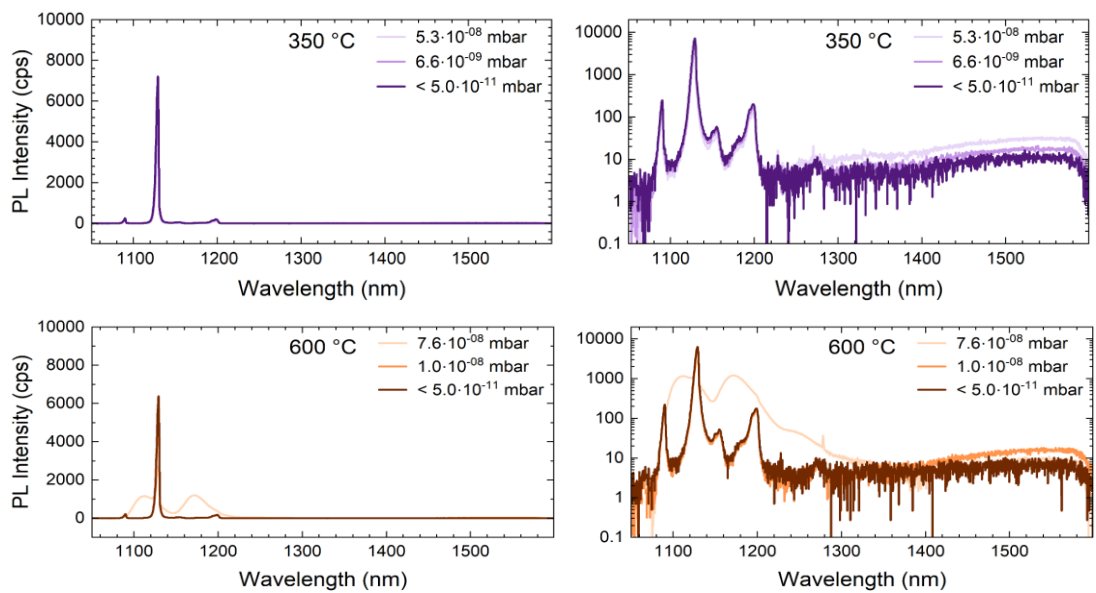

**Figure S1. PL spectra of ULT-grown Si, deposited at 350°C and 600°C for different growth pressures.** For PL spectra, plotted on a linear scale there seems to be no PL signal below the Si-related bulk phonon replica, thus, the spectra appear to relate to defect free samples. Semi-logarithmic spectra can reveal PL defect backgrounds that indicate changes in the matrix quality.

## W-Centers

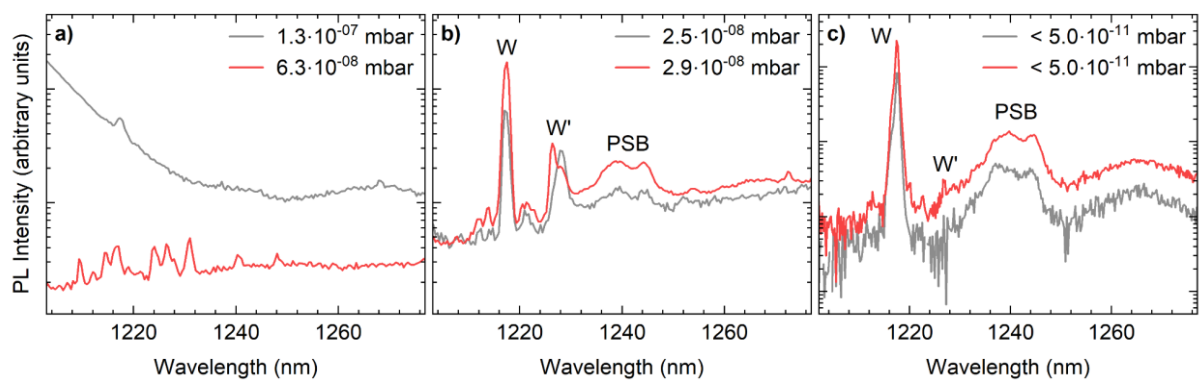

**Figure S2.** PL spectra of W-center samples grown at different growth pressures following the sample fabrication schemes in Figure 1b,c. Red spectra: A 9 nm thick Si layer was grown at $T_G$ = 200°C, and capped at $T_{cap}$ = 250°C. These spectra are compared to its reference spectra (grey spectra) for which the ULT growth was conducted at $T_{cap}$ = 250°C throughout (Figure 1c).

Figure S2 shows the influence of the growth pressure on the PL emission of W-centers. The spectra shown in red have been recorded from samples following the fabrication scheme depicted in Figure 1b of the main text, i.e., a 9 nm thick layer of Si was deposited at 200°C, followed by the growth of a 105 nm thick Si capping layer, grown at 250°C. These spectra are compared to the grey reference spectra in Figure S2, which show the PL response of samples that followed the sample fabrication scheme in Figure 1c of the main text, i.e., the ULT-grown layers were grown at 250°C throughout. For the worst pressure conditions (Figure S2a), no pronounced signal from W-centers is visible. The reference signal is dominated by a broad PL background that is significantly reduced when the 9 nm thick Si layer grown at 200°C is added (red spectrum in Figure S2a), for which, in turn, only faint lines around 1220 nm are visible. Due to the poor growth environment, it is not possible to attribute these lines to specific, known defects. For the growth at pressures that are dominated by a high $H_2$ partial pressure (Figure S2b), a clear W and W'-center ZPL is visible. For the best investigated pressure (Figure S2c) the spectra are dominated by W-center ZPL. We note that, under the employed growth conditions, the vertical position control for W-centers is poor, as can be seen in the intensity ratio of the W-center ZPL between the red and grey spectra in Figure S2c. We suggest that a refined temperature ratio between the active layer and capping layer is necessary to enhance the vertical creation selectivity using this MBE self-assembly approach.